\RequirePackage{fix-cm}
\documentclass{article}
\usepackage{graphicx}
\usepackage[]{amsmath}
\usepackage{chemist}
\newcommand{\varA}[1]{{\operatorname{#1}}}

\begin{document}

\title{A Novel Simplified Model for Blood Coagulation: A piecewise dynamical model for thrombin with robust predictive capabilities}

%



\author{Jayavel Arumugam\thanks{jayavel.arumugam.17@gmail.com, Dept. of Mechanical Eng., Texas A \& M University}        \and
        Arun Srinivasa\thanks{arun.r.srinivasa@gmail.com, Dept. of Mechanical Eng., Texas A \& M University}
}



\date{\today}

\maketitle

\begin{abstract}

Realistic description of patient-specific mechanical properties of clotting dynamics presents a major challenge. Available patient-specific data falls short of robustly characterizing myriads of complex dynamic interactions that happen during clotting. We propose a simplified switching model for a key part of the coagulation cascade that describes dynamics of just four variables. The model correctly predicts prolonged activity of thrombin, an important enzyme in the clotting process, in certain plasma factor compositions. The activity sustains beyond the time which is conventionally considered to be the end of clotting. This observation along with the simplified model is hypothesized as a necessary step towards effectively studying patient-specific properties of clotting dynamics in realistic geometries. 

\end{abstract}


\section{Introduction}
\label{intro}


Patient-specific geometry modeling, simulations with more realistic boundary conditions, multiscale models that combine molecular mechanisms with clinical manifestation are some of the open problems discussed in vascular biomechanics \cite{taylor2009open}. Blood constituents vary drastically in patients and their composition plays a significant role in determining the mechanical properties of the resulting clot \cite{undas2014fibrin,van2016constitutive,wolberg2007thrombin}. 

Current methods model generation and depletion of various constituents in blood flow using convection reaction diffusion equations \cite{rajagopal2003review},

\begin{equation}
\frac{\partial [Y_{i}]}{\partial t}
+  
div([Y_{i}] \mathbf{v})
= 
div(D_{i} grad [Y_{i}]) 
+
G_{i},
\quad 
i = 1, ..., N, 
\label{eq:CRD}
\end{equation}
where $[Y_{i}]$ is a constituent, $\mathbf{v}$ is velocity, $D_{i}$ are diffusion coefficients, and $G_{i}$ are source terms due to clotting reaction dynamics. Number of constituents $N$ is in the order of thousands; typical models consider between 30 to 40 constituents \cite{hockin2002model,rajagopal2003review}; and clotting reaction models considering hundreds of constituents have been proposed \cite{luan2007computationally,luan2010ensembles}. 

\subsection{Need for Simple Models}
Models accounting for myriads of chemical reactions are considered as a first step towards understanding the complex phenomena of clotting \cite{rajagopal2003review}. Further modeling many such constituents are useful in a diagnostic setting where often diseases are reflected as abnormalities in the coagulation cascade \cite{rapaport1993blood}. In particular reaction dynamics alone, for example, dynamics of thrombin \cite{mann2003dynamics,brummel2013riskdisease}, is known to be associated with many diseases such as acute coronary syndrome \cite{brummel2008ACDvsCAD}, acute cerebrovascular disease \cite{gissel2010plasma}, rheumatoid arthritis \cite{undas2010thrombin}, etc.  Recent studies in this direction include use of machine learning tools to characterize and use such abnormalities for disease classification tasks such as identifying patients prone to heart attacks \cite{arumugam2016random,jayavel2017dissertation}. 

However, simplified models for the coagulation cascade are required so that:
\begin{enumerate}
\item the model predictions are more readily verifiable using experiments  
\item augmenting chemical reaction dynamics in other phenomena such as flow and mechanical characterization of clots in patient-specific terms become feasible and readily comprehensible. 
\end{enumerate}


\subsection{Challenges in Modeling the Coagulation Cascade}
\label{ssec:challeneges} 

The chemical reaction kinetics particularly poses many challenges. Models describing kinetics using $M$ reactions are of the form,

\begin{align}
\frac{d [Y_{i}]}{d t} 
= G_{i} 
&= 
\sum_{j=1}^{M} S_{ij} r_{j} ([\mathbf{Y}]; \mathbf{k}_j ), \notag \\
s.t. \quad [Y_{i}](0) & = [Y_{i}^{0}], \notag \\
\mathbf{S}[\mathbf{Y}] &= \mathbf{S}[\mathbf{Y}_{}^{0}], \notag \\
[Y_{i}] &\geq 0, 
\end{align}

\noindent where $S_{ij}$ are stoichiometric coefficients ($ i = 1, ..., N, j = 1, ..., M $), $r_{j}$ are reaction rates,  $\mathbf{k}_j$ are parameters of the reaction rates, and $[Y_{i}^{0}]$ are the initial plasma composition. 

\begin{figure}[htbp]
\centering
\includegraphics[width=0.8 \textwidth]{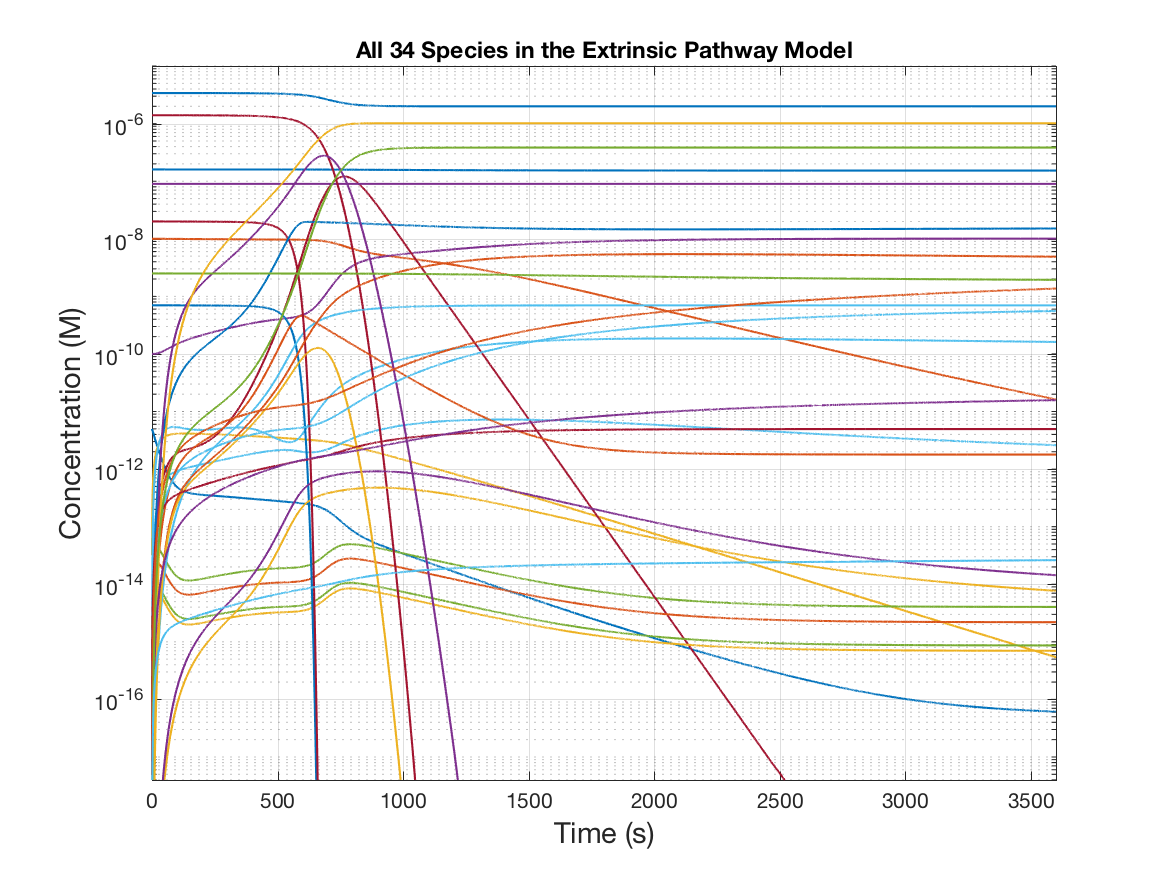}
\caption[Concentration of different species during clotting in the extrinsic pathway]{Concentration of different species during clotting in the extrinsic pathway. The concentration of various species varies by orders of magnitude. This poses stiffness issues in numerical schemes demanding small time steps and excessive conditioning.} 
\label{fig:concOrders}
\end{figure}

Concentration of proteins involved in coagulation and rates of reactions vary by orders of magnitude (see Figure \ref{fig:concOrders}). Essentially, picomoles of trigger results in the formation of hundreds of nanomoles of certain enzymes. This in turn results in macroscopic formation of clots. Typically stochastic methods \cite{gillespie1977exact} are used to properly account for low concentrations of species \cite{laurenzi1999monte,lo2006stochastic}.

The nonlinear chemical kinetics is modeled using reaction rates that typically has quadratic terms or Hill-like terms \cite{weiss1997hill} .  The models are very stiff and solution trajectories are unstable in many directions \cite{danforth2009impact}. The rates involve negative feedback loop or cycles in the reaction cascade \cite{ataullakhanov2005mathematicalNegFeed}. This poses challenges in numerically coupling chemical reactions with flow simulations.

Moreover, there is uncertainty in the parameters of the reaction kinetics models. It is hard to measure most of these protein factors. Many rate parameters are inferred indirectly rather than being directly measured.
Models used to simulate clot in small two dimensional regions ($\sim$100 $\mu$m) consider the dynamics of many reactants \cite{papadopoulos2014simplified}. These are inappropriate for simulations in realistic 3 dimensional flow conditions in arteries, say, in order to study atheroscelerosis or thrombosis \cite{papadopoulos2014simplified}. Reduced dimensional simplified chemical kinetics models will help to advance the quality of patient specific simulations in realistic geometries. 

\subsection{Relevant Literature on Reduced Order Models} 
Papadopoulos et al. \cite{papadopoulos2014simplified} suggested a phenomenological model for thrombin dynamics. 
Based on the mechanism of thrombin dynamics, they propose a simplified thrombin dynamics model using four reactions. The reactions include dynamics of thrombin, prothrombin, platelets, and activated platelets. Using the assumption of fast platelet activation, they derive analytical expression for thrombin generation. These are similar to thrombin generation functions prescribed by Hemker et al. \cite{wagenvoord2006limits}. The model essentially fits patient specific thrombin generation profiles and the effect of the plasma factor composition and inhibitors on the dynamics of thrombin were not emphasized. 

This motivated Sagar et al. \cite{sagar2015dynamic} to come up with a dynamical model for thrombin generation using a hybrid strategy. The strategy combines differential equations and several logical rules to model thrombin dynamics. They design their approach to model systems where mechanistic insights are poor and experimental interrogation is difficult. This results in reduced order model that has rates of the product of Hill-like terms and transfer functions that act as the logical rules, i.e., 

\begin{equation}
 r_{i} = \frac{k_{i} x_{i}^{\eta_{i}} }{ 1 + k_{i} x_{i}^{\eta_{i}} } 
\text{min}\left( \frac{k_{j} x_{j}^{\eta_{j}} }{ 1 + k_{j} x_{j}^{\eta_{j}} }, \frac{k_{m} x_{m}^{\eta_{m}} }{ 1 + k_{m} x_{m}^{\eta_{m}} } \right)
\label{eq:hillRate}
\end{equation}

\noindent where $r$ is the reaction rate, $k$ and $\eta$ are parameters, and $x$ pertains to protein concentration or activity. Though the model shows good performance for thrombin dynamics, the transparency in the mechanistic models such as \cite{wagenvoord2006limits} and  \cite{papadopoulos2014simplified} is lost, i.e., the functionality of the model parameters and effect of their changes is not evident. We seek a middle ground between the two simplified models where we find a dynamical model that makes use of the mechanistic knowledge of blood coagulation and is able to account for changes in plasma factor composition.  

\subsection{Proposed Model Properties}
We suggest a simple phenomenological model for thrombin dynamics based on chemical kinetics:  
\begin{enumerate}
\item We model the stoichiometry and dynamics of certain important and easily measurable chemical species. The model is based on the classically viewed initiation, propagation, and termination of thrombin dynamics. Hence the chemistry involved in the simulations offer physiological insight. 
\item We model the initiation phase and the propagation/termination phase of thrombin dynamics separately using a switching criteria for the rate parameters. The switch separates the slow initiation phase from the latter which is orders of magnitude faster. 
\item  The model we propose shows varied responses. The functionality of the parameters are evident and different aspects of thrombin dynamics are easily alterable. 
\item A good model should be able to capture the necessary rich behavior of the phenomena as well as generalize well in order to predict important qualitative and quantitative responses. The model accounts for the physiological effect of antithrombin and is able to predict certain important changes in thrombin dynamics due to changes in prothrombin and antithrombin concentration. 
\item The model is dynamical in the sense that rates for the model species could be calculated given the current values of the model variables. In other words, we essentially model $G_{i}$ for the important species in equation \ref{eq:CRD} so that the model could be extended to account for transport.  
\end{enumerate} 

The simplified model proposed here is expected to make patient-specific mechanical characterization of clots in realistic geometries feasible.

\begin{figure}[htbp]
\centering
\includegraphics[width=0.8 \textwidth]{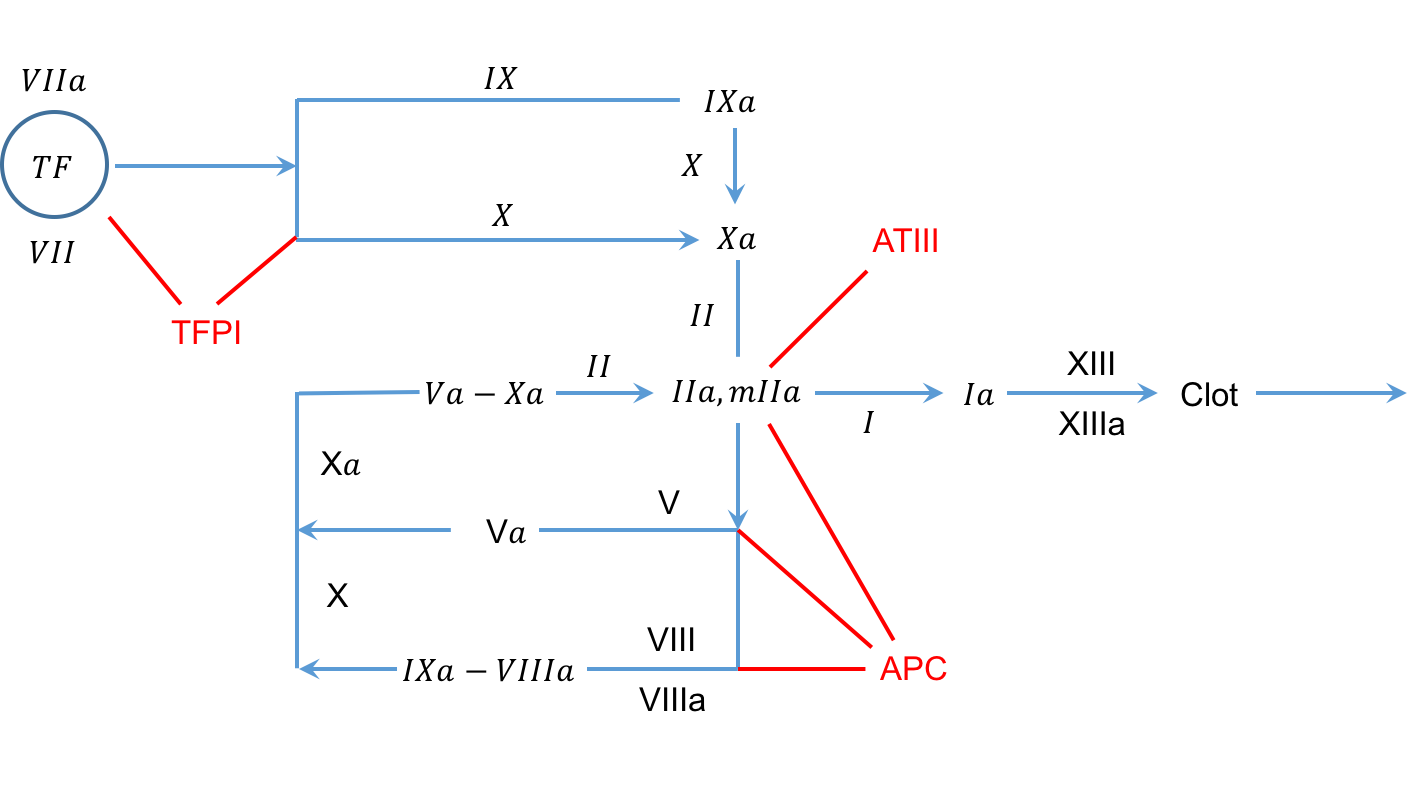}
\caption[A schematic of the extrinsic pathway]{A schematic of the extrinsic pathway. We propose a simplified model for thrombin dynamics. We note that events that occur after thrombin generation result in changes of mechanical properties.} 
\label{fig:ExtrinsicnIntrinsic6}
\end{figure} 

\section{Thrombin Generation in the Extrinsic Pathway}
\subsection{Full Model}
\begin{figure}[htbp]
\centering
\includegraphics[width=0.8 \textwidth]{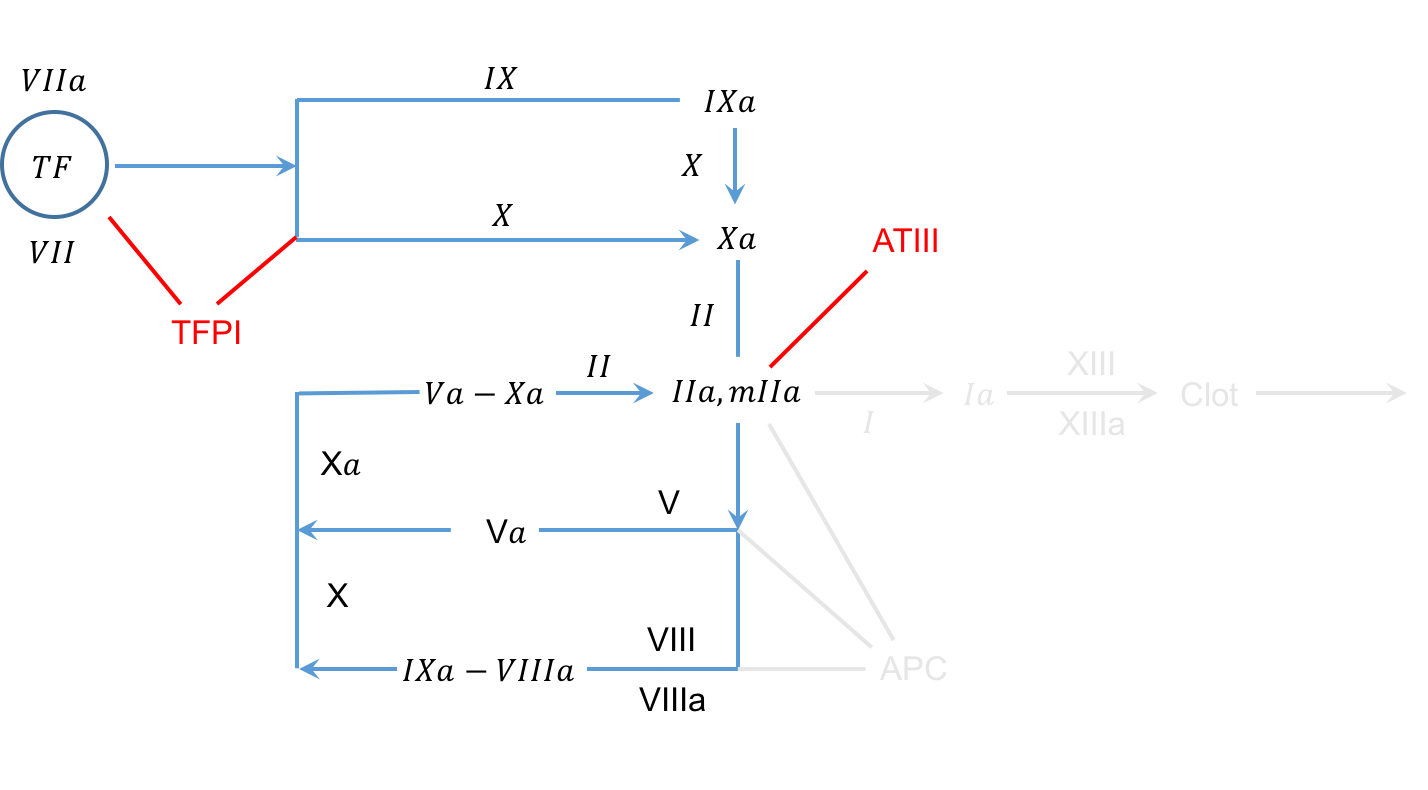}
\caption[A schematic of the extrinsic pathway model for thrombin dynamics used in this work]{A schematic of the extrinsic pathway model for thrombin dynamics used in this work. There are essentially three elements in this network, namely, i) thrombin initiation; ii) thrombin propagation; and iii) thrombin inhibition.} 
\label{fig:ExtrinsicMann6}
\end{figure}

Figure \ref{fig:ExtrinsicnIntrinsic6} shows a schematic of the key elements of the extrinsic pathway involved during clotting. We consider the extrinsic pathway because hemostasis occurs due to tissue factor initiation. Further, we simply simplify thrombin dynamics. Given that flow properties affect and are affected by fibrin formation, such a simple thrombin dynamics model factors out the two phenomena. Reactions in the intrinsic pathway could be readily accounted for due to the phenomenological aspect of initiation in the proposed model. For simulations of the full model, we use the extrinsic pathway developed by Hockin et al. \cite{hockin2002model}. A schematic of the model used is shown in Figure \ref{fig:ExtrinsicMann6}. 

\subsection{Key Ideas Used for Simplification}
We exploit the following ideas for model simplification:

\paragraph{Dynamical Parametrization:} Patient specificity in blood coagulation is usually modeled using variations in the initial conditions for the differential equations. This leads to a high-dimensional state space and specification of dynamics in that space of many constituents as described earlier. However, we simplify the state of the system using just four important species and then model the dynamics of those four constituents. The effect of other constituents are parametrized in the rates describing the dynamics of the former, i.e., the effect of patient specificity of thrombin dynamics is described using variations in the rate parameters of the simplified model. We show that training the model for a specific choice of initial condition in the original state space (physiological mean) is able to predict qualitative responses of changes in prothrombin and antithrombin concentration in the reduced state space. 

\paragraph{Switching:} We separate the parameters in the initiation and propagation/termination phase of thrombin dynamics using a switching model. Such switching models are commonly used to  model complex nonlinear systems \cite{fox2009bayesian}. Switching models are known to result in simplified models for even chaotic dynamics. For example, piecewise affine models are known to model chaotic behavior of many nonlinear systems well \cite{amaral2006piecewise}. The proposed model switches from initiation to propagation/termination phase based on the concentration of thrombin. Such threshold-based responses have been used in models describing initiation \cite{papadopoulos2014simplified} and platelet activation \cite{rajagopal2003review} in the coagulation cascade. The hybrid (switching) model by Makin and Narayanan \cite{makin2013hybrid} intends to develop a comprehensive model rather than a simplified model.


\begin{figure}[htbp]
\centering
\includegraphics[width=0.6 \textwidth]{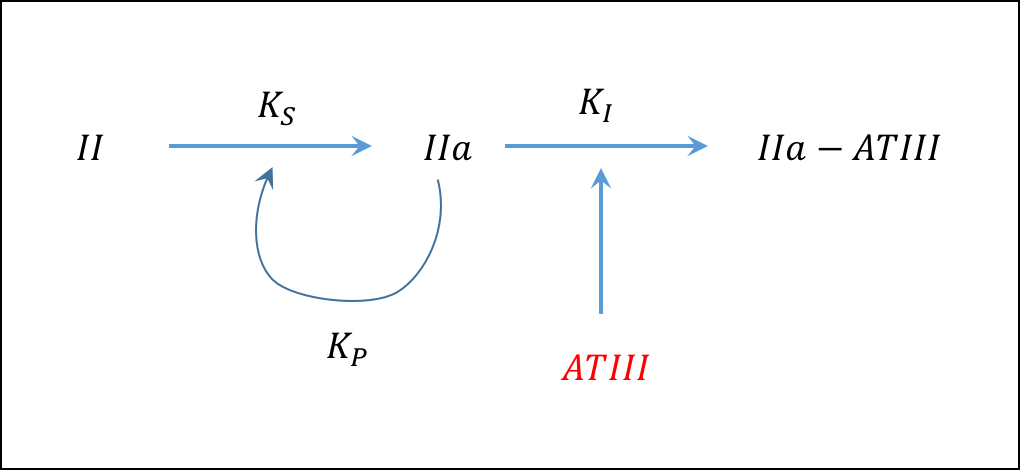}
\caption[A schematic of the simplified model proposed in this study]{A schematic of the simplified model proposed in this study. There is fuel prothrombin, the bursting enzyme thrombin, the inhibitor antithrombin, and the by-product thrombin-antithrombin. $K_S$ is a rate constant that models initiation due to injury. $K_P$ is the rate of thrombin propagation. $K_I$ is the rate of thrombin inhibition.} 
\label{fig:ExtrinsicJay}
\end{figure} 

\subsection{Simplified Model}
We use the traditional simplification of the thrombin generation cascade and describe kinetics for prothrombin, thrombin, antithrombin, and thrombin-antithrombin using the following set of reactions: 
\begin{enumerate}
\item \textbf{Thrombin Initiation}: Tissue factor activates prothrombin to form thrombin. 

\begin{chemmath}
II \reactrarrow{0pt}{2.5cm}{$K_{S}$}{Initiation} IIa
\end{chemmath}

\item \textbf{Thrombin propagation}: Given that sufficient amount of thrombin (2 nM) is activated, clotting propagates via a different set of reactions. 

\begin{chemmath}
II \reactrarrow{0pt}{2.5cm}{$K_{P}$}{Propagation} IIa
\end{chemmath}

\item \textbf{Thrombin inhibition}: Finally normal hemostasis requires that thrombin generation is controlled. 

\begin{chemmath}
IIa + AT \reactrarrow{0pt}{2.5cm}{$K_{I}$}{Termination} IIaAT
\end{chemmath}

\end{enumerate}

\begin{figure}[htbp]
\centering 
\includegraphics[width=0.8 \textwidth]{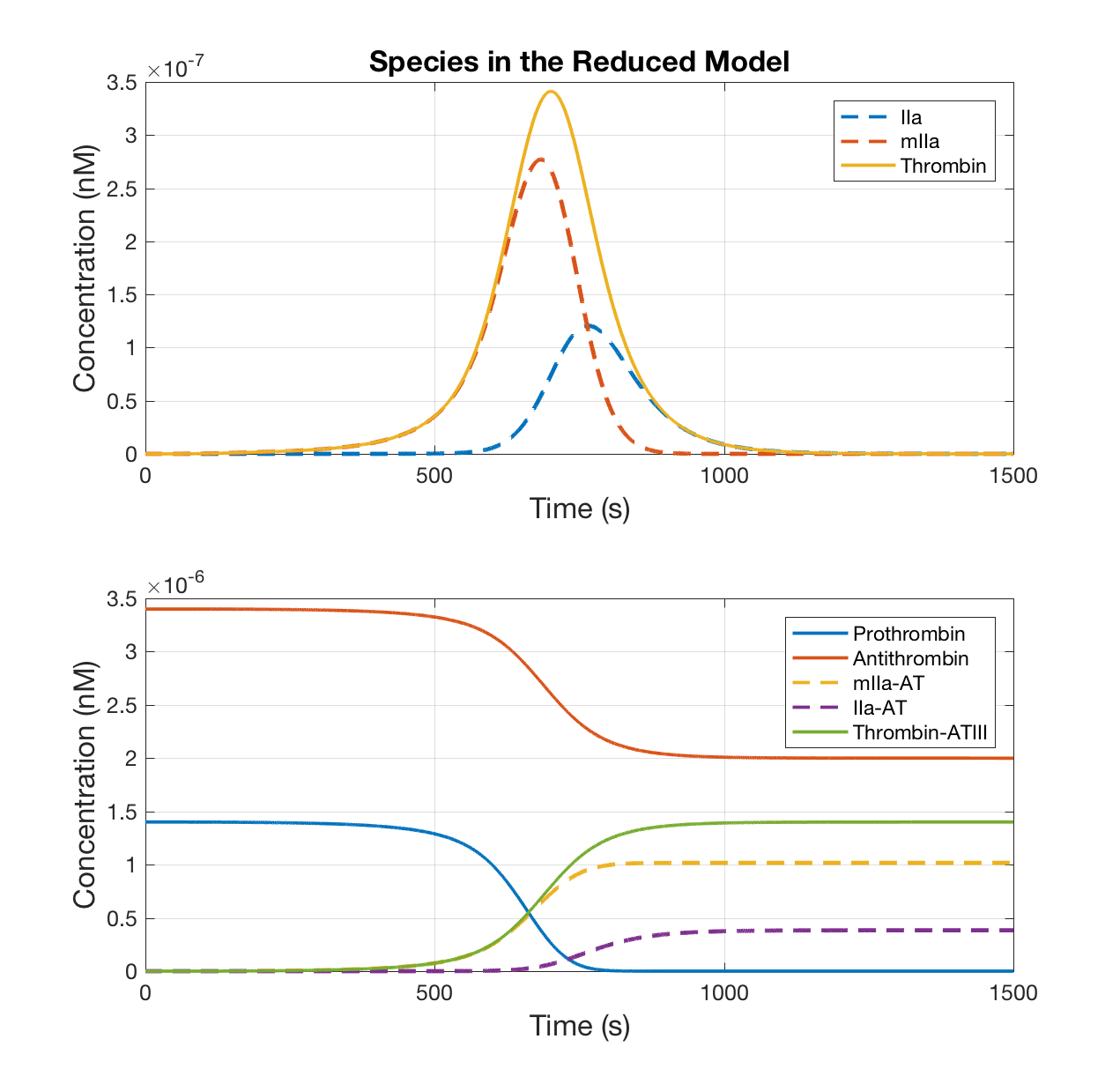}
\caption[Stoichiometry of thrombin and antithrombin]{Stoichiometry of thrombin and antithrombin. Only the species drawn with a continuous straight line are considered in the model. This ensures stoichiometric consistency. [IIa] or thrombin in our model is sum of the two forms of thrombin in the full model. Similarly, [IIa-AT] or thrombin-antithrombin in our model the sum of the two by products of thrombin inhibition due to antithrombin in the full model.} 
\label{fig:ThrombinGenStoichiometry}
\end{figure}

Corresponding to the reactions, we define the following stoichiometric model: 
\begin{align}
&\frac{d}{dt} \varA{[II]} = - K_{S} - K_{P}  \varA{[II]} \varA{[IIa]}   \notag \\
&\frac{d}{dt} \varA{[IIa]} = K_{S} + K_{P}  \varA{[II]} \varA{[IIa]}   -  K_{I}  \varA{[IIa]}  \varA{[AT]} \notag \\
&\frac{d}{dt} \varA{[AT]} = - K_{I}  \varA{[IIa]}  \varA{[AT]} \notag \\
&\frac{d}{dt} \varA{[IIa-AT]} = K_{I}  \varA{[IIa]}  \varA{[AT]}.  
\label{eq:simplfiedModelReactionsRates}
\end{align}
\noindent where we model initiation using the rate constant $K_S$, propagation using the rate constant $K_P$, and inhibition using the rate constant $K_{I}$. In order to be stoichiometrically consistent, our [IIa] is the sum of both forms of thrombin in the full 34 variable model (the variables used in our model are drawn in continuos lines in the full model simulation as seen in Figure \ref{fig:ThrombinGenStoichiometry}). [IIa-AT] in our model is the sum of the two antithrombin complex formed due to inhibition\footnote{there are other antithrombin complexes formed in the full model but they are 3 orders of magnitude smaller than [IIa-AT] and we neglect them.}. 

Among the four species modeled, methods to measure or infer thrombin, ATIII, and thrombin antithrombin  \cite{abildgaard1977antithrombin,pelzer1988determination,macfarlane1953thrombin,hemker2003calibrated} exist. 
The rates in equation \ref{eq:simplfiedModelReactionsRates} are such that they satisfy a stoichiometric constraint, i.e., 

\begin{align}
\varA{[II]} + \varA{[IIa]} + \varA{[IIa-AT]} &= c_{1}  \notag \\
\varA{[AT]} + \varA{[IIa-AT]} &= c_{2}  
\label{eq:stoichiometricInvariance}
\end{align}
\noindent $c_{1}$ and $c_{2}$ are constants defined using the initial conditions. These dependence relations are crucial while designing experiments to properly observe the system. For example, calibrated automated thrombogram assay is inadequate to make proper measurements with deficient ATIII \cite{knappe2015application} due to fluorogenic substrate depletion. In the results section, we will discuss the possibility of sustained thrombin activity when ATIII is deficient leading to the depletion.  


\begin{table}
\caption{Switching Condition}
\label{tab:switchingRules}       
\centering
\begin{tabular}{lll}
\hline\noalign{\smallskip}
 & [IIa] $ < $ 2 nM & [IIa] $ \geq $ 2 nM  \\
\noalign{\smallskip}\hline\noalign{\smallskip}
$K_{S}$ & $k_{s} > 0$ & 0 \\
$K_{I}$ & $k_{i2} > 0$ & $k_{i1} > 0$ \\
$K_{P}$ & 0 & $k_{p} > 0$ \\
\noalign{\smallskip}\hline
\end{tabular}
\end{table}

We propose the switching rules in Table \ref{tab:switchingRules} that changes the response of the model during initiation and propagation/termination. Essentially, thrombin propagation occurs if [IIa] crosses a threshold. For normal clotting, rate of propagation is expected to be orders of magnitude higher than that of rate of initiation. We also use two different inhibition rate constant $k_{i1}$ and $k_{i2}$ so that rate of inhibition could be separately tuned during the two phases.

\section{Results and Discussion}
\subsection{Estimation of Model Parameters}
In the simulations of the full model, clotting was initiated with 5 pM and the plasma factor composition was set to physiological mean values \cite{hockin2002model}. We used the data from the full model to fit parameters for the reduced model. We note that thrombin dynamics of the full model has been corroborated with experimental data \cite{hockin2002model}. Particle swarm optimization \cite{eberhart1995new} was used for parameter estimation and we obtained one set of parameters for the physiological mean composition.

\begin{figure}[htbp]
\centering
\includegraphics[width=0.8 \textwidth]{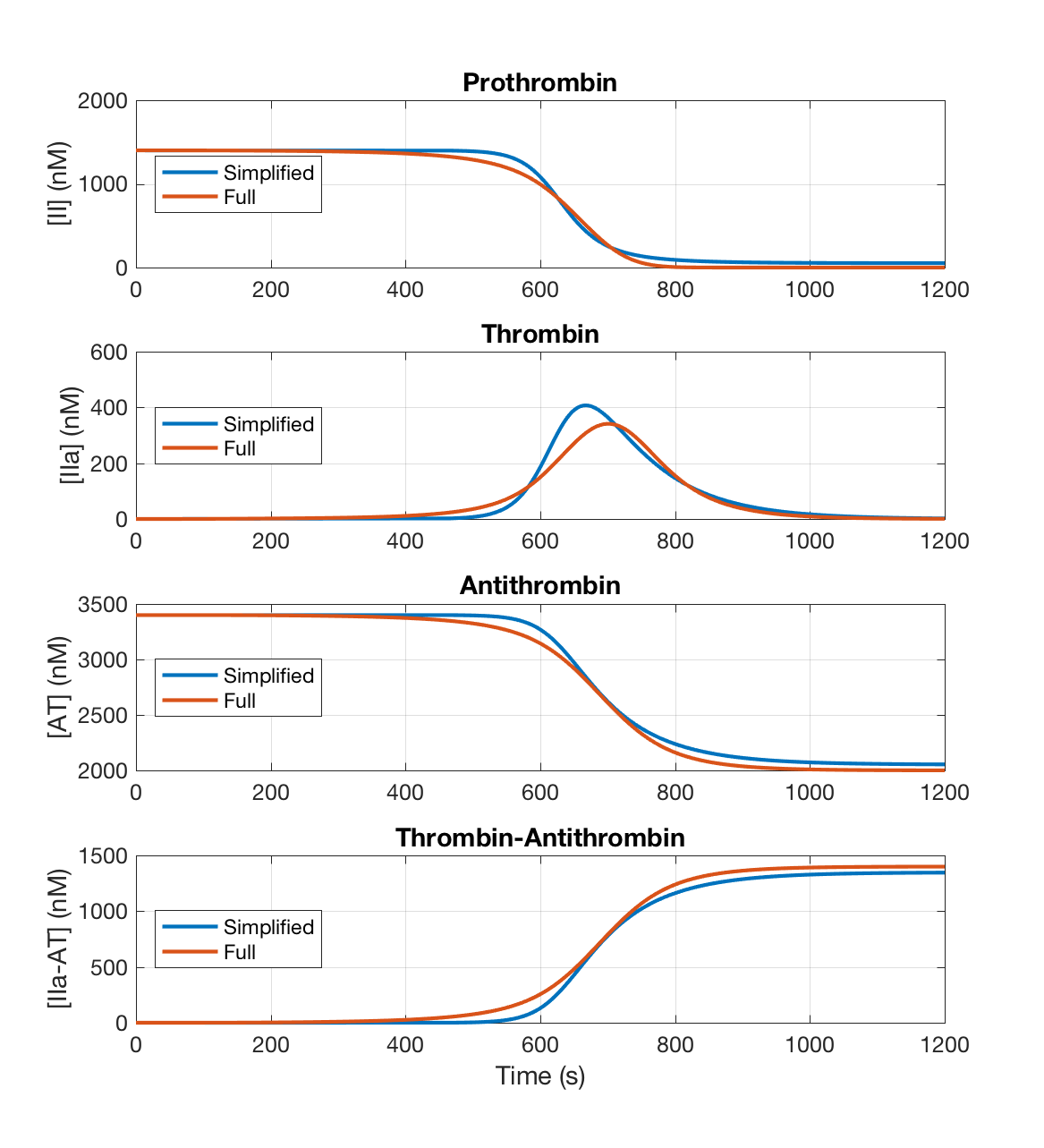}
\caption[Comparison of all the species modeled in the simplified model]{Comparison of all the species modeled in the simplified model. The simplified model captures the reponse of the corresponding variable in the full model very well. There is a slight mismatch in the maximum amount of thrombin generation. This could be improved by also choosing the clot propagation threshold ([IIa] = 2 nM) better. } 
\label{fig:ThrombinGen4SpeciesComparison}
\end{figure} 

We used the sum of squared differences of the normalized concentration profiles between the common species in the full model and the reduced model as the objective function,
\begin{equation}
u = \frac{1}{M} \sum_{m=1}^{M} \sum_{i=1}^{4} \left( \frac{ C_{i}^{reduced}(m;k)  - C_{i}^{full}(m)}{C_{i}^{constant}} \right)^{2}
\label{eq:simplificationObjective}
\end{equation}
\noindent where $m$ denotes time points and $i$ denotes model species. Mean concentration of prothrombin and antithrombin were used as normalization constants $C_{i}^{constant}$. Comparison of the reduced and full model simulation for the physiological mean initial composition is shown in Figure \ref{fig:ThrombinGen4SpeciesComparison}.

\subsection{Parameter Study of the Simplified Model}
\begin{figure}[htbp]
\centering
\includegraphics[width=0.6 \textwidth]{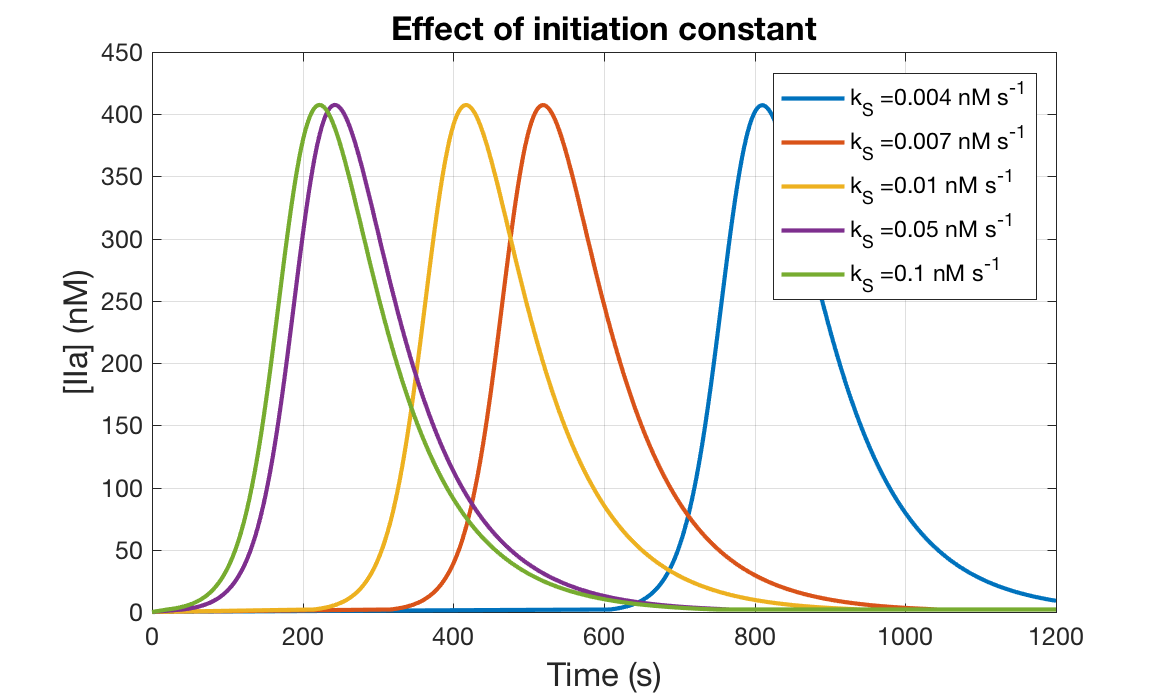}
\caption[Controlling thrombin initiation using $K_S$]{Controlling thrombin initiation using $K_S$. There is an nonlinear dependence of clot time on this parameter due to $K_{i2}$. } 
\label{fig:EffectofKStart}
\end{figure}

\begin{figure}[htbp]
\centering
\includegraphics[width=0.6 \textwidth]{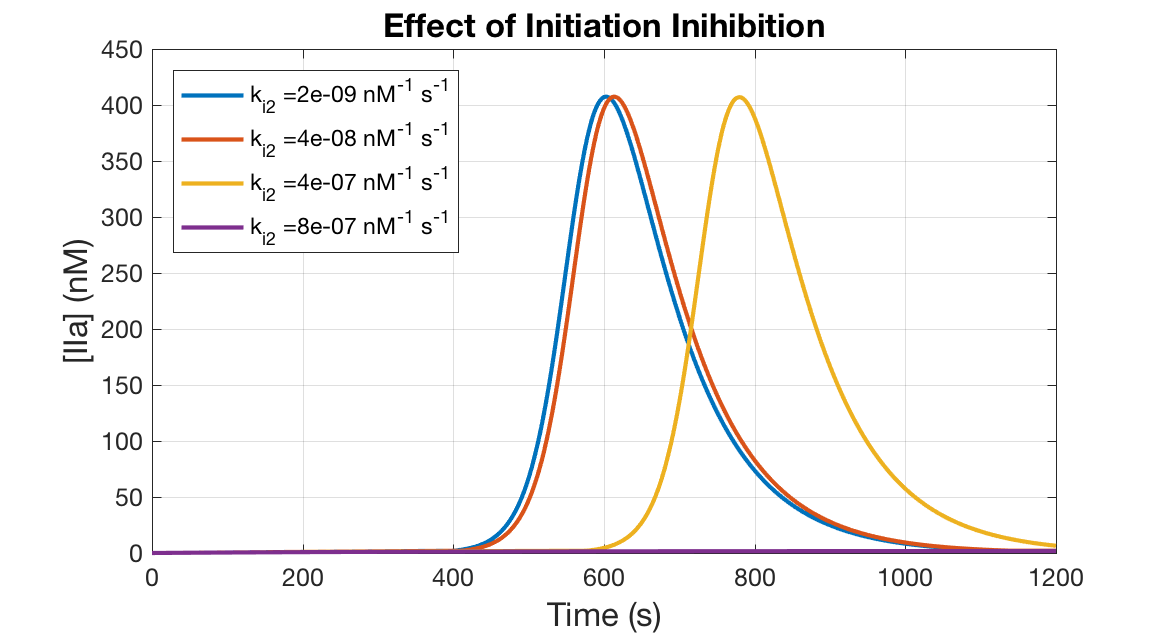}
\caption[Controlling thrombin initiation using $k_{i2}$]{Controlling thrombin initiation using $k_{i2}$. This parameter along with $K_S$, allows for modeling a wide range of clot times and dynamics during initiation.} 
\label{fig:EffectofKStart_inhibition}
\end{figure}

We show the effect of parameters by changing them one at a time. Clot time depends exponentially on $K_S$ (seen in Figure \ref{fig:EffectofKStart}). Rate constant $k_{i2}$ also controls the clot time Figure \ref{fig:EffectofKStart_inhibition}. For certain combinations of $K_S$ and $k_{i2}$ it takes more than 1200 seconds for clot initiation. Both the parameters together offer more control over dynamics of clot initiation. 

\begin{figure}[htbp]
\centering 
\includegraphics[width=0.6 \textwidth]{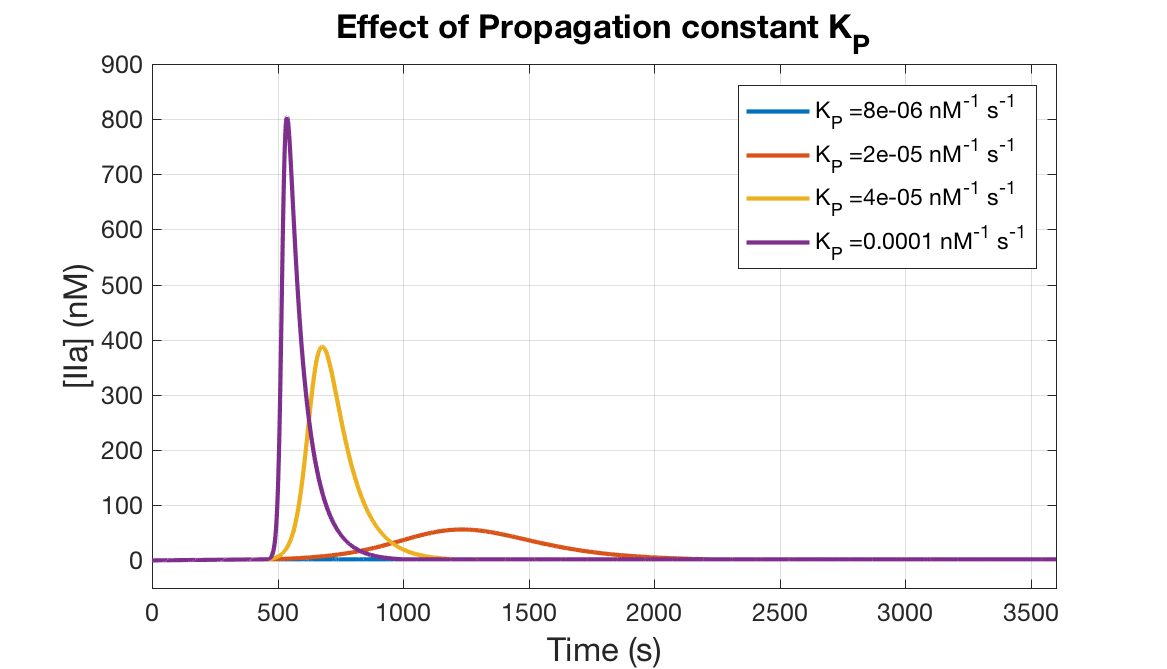}
\caption[Controlling thrombin propagation using $K_P$]{Controlling thrombin propagation using $K_P$. As expected, variations in the propagation rate constant is able to capture a wide range of thrombin generation rates.}  
\label{fig:EffectofKP_propagation}
\end{figure} 

\begin{figure}[htbp]
\centering
\includegraphics[width=0.6 \textwidth]{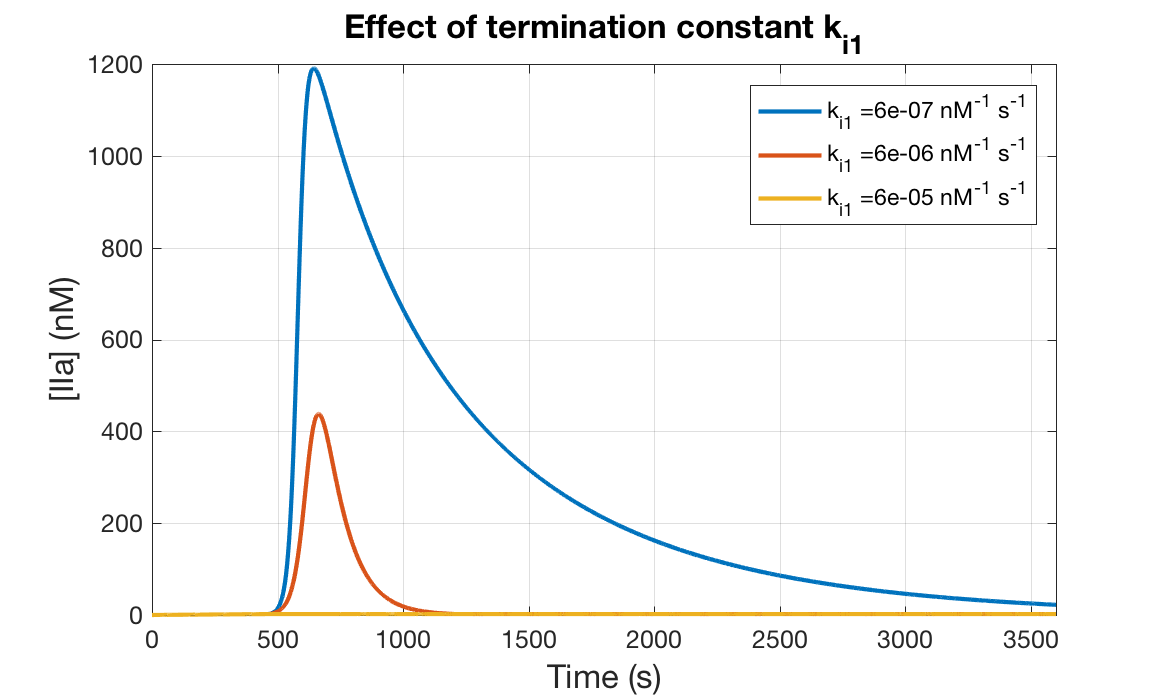}
\caption[Controlling thrombin termination using $k_{i1}$]{Controlling thrombin termination using $k_{i1}$. This parameter has more effect on the termination phase of thrombin generation.} 
\label{fig:EffectofKI_termination}
\end{figure}

Figures \ref{fig:EffectofKP_propagation} and  \ref{fig:EffectofKI_termination} show the effect of changes in the rate constants $K_P$ and $k_{i1}$ respectively. The parameters offer a wide range of thrombin generation rates. Similar to the initiation, there are certain values of $K_P$ (for a given value of $k_{i1}$) and vice versa where thrombin generation is too low. These two parameters together offer control over simulating a wide range of thrombin propagation.

\subsection{Prediction of Variation in Prothrombin and Antithrombin}
\begin{figure}[htbp]
\centering
\includegraphics[width=0.6 \textwidth]{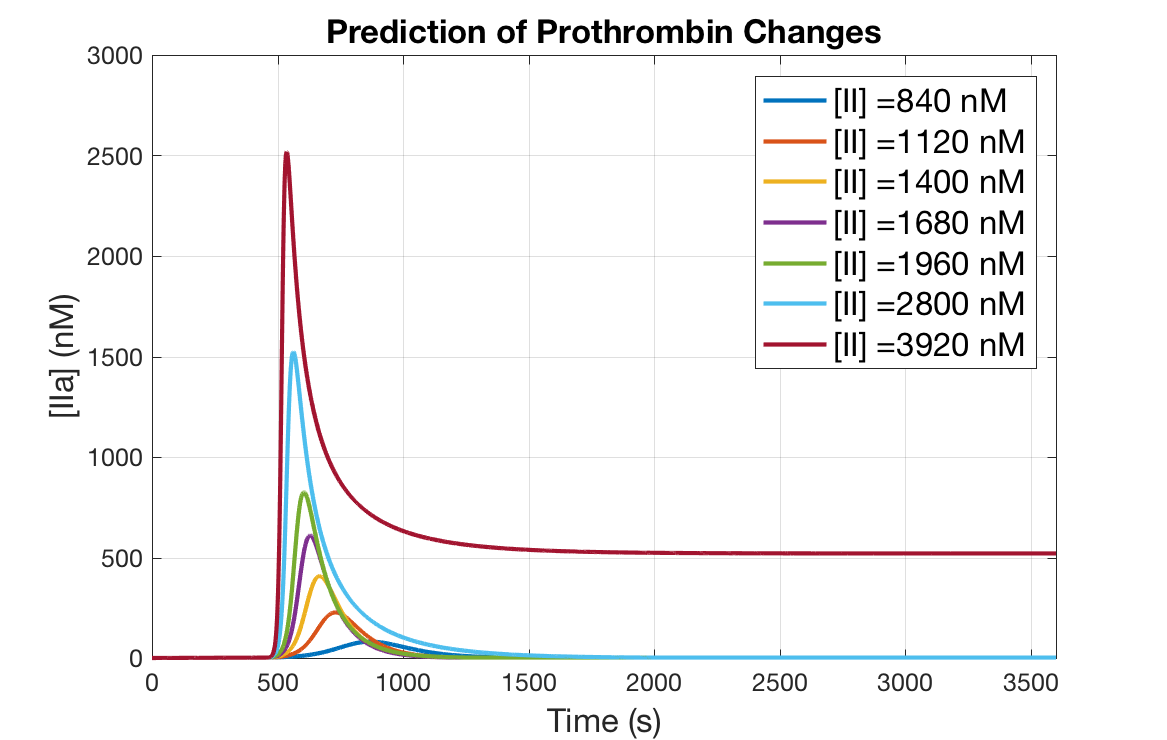}
\caption[Prediction on prothrombin variation]{Prediction on prothrombin variation. Thrombin activity sustains in plasma factor composition that has excessive prothrombin. Similar prediction is made in Figure \ref{fig:prediction_antithrombin} based on variations in initial antithrombin concentration.} 
\label{fig:prediction_prothrombin}
\end{figure}

Finally, we check the qualitative response of the model predictions for variations in initial prothrombin and antithrombin concentrations. As seen in Figure \ref{fig:prediction_prothrombin}, higher values of prothrombin are able to predict more thrombin generation. This has been observed in experiments \cite{allen2004impact}. Thrombin rates during termination could be improved using better training data and using different reaction rates. Similarly, lower values of antithrombin are able to predict higher thrombin generation. 

\begin{figure}[htbp]
\centering
\includegraphics[width=0.6 \textwidth]{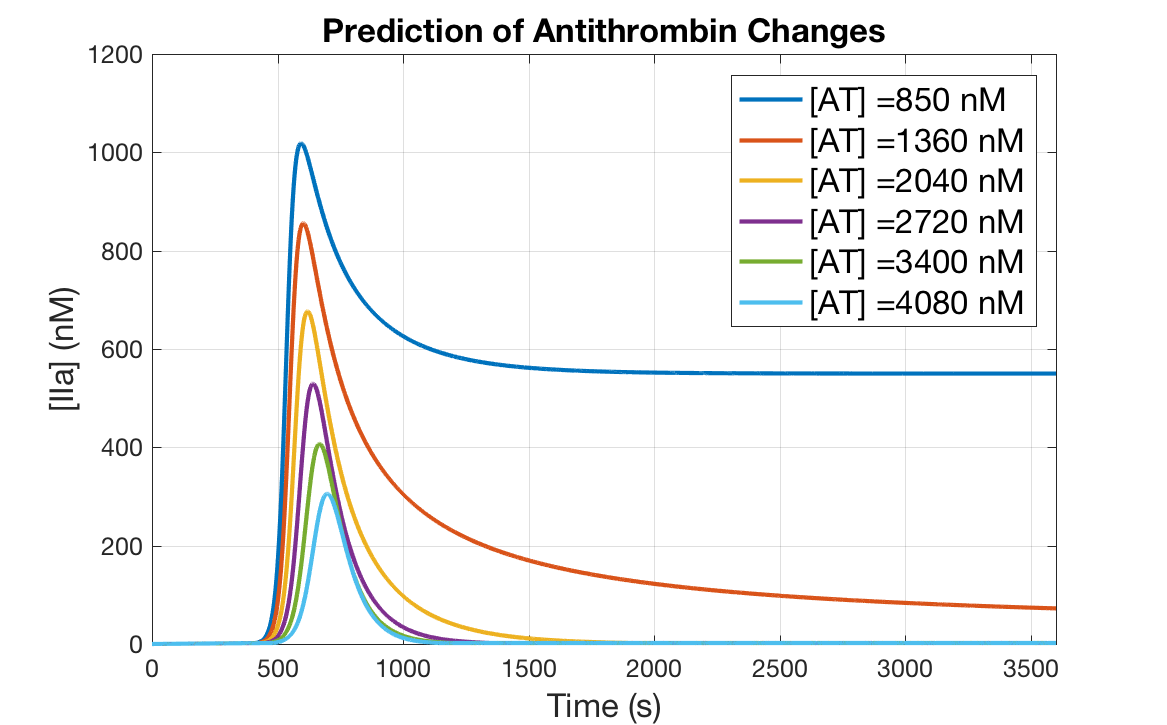}
\caption[Prediction on antithrombin variation]{Prediction on antithrombin variation. This is the most important and significant prediction of the model. This has been observed in thrombin generation experiments \cite{allen2004impact}. Moreover, this phenomena could be blind to markers like TAT (IIa-AT) that infer thrombin activity.}  
\label{fig:prediction_antithrombin}
\end{figure}

One of the most important predictions of this model is that thrombin termination appears to halt at non-zero values (Figure \ref{fig:prediction_antithrombin}). Such sustained activity is also in observed in experiments when there is too much prothrombin compared to antithrombin \cite{allen2004impact}. Inhibitors such as activated protein C may need to modeled in order to account for oscillations observed in such sustained activity. In this model, the reaction essentially runs out of the inhibitor [AT] when initiated with a certain plasma factor composition. In such a scenario, other phenomena like diffusion and convection will control the extent of clotting. For example, when there is less inflow of antithrombin concentration, as in the case of stasis, we would expect more clotting due to the presence of excess active thrombin. The significance of antithrombin deficiency in simulations accounting for diffusion has already been reported \cite{anand2008model}. For a given value of initial prothrombin, and given that thrombin generation proceeds normally due to sufficient rates, the simplified model quantifies the amount of antithrombin required for thrombin generation to terminate.

\begin{figure}[htbp]
\centering
\includegraphics[width=0.6 \textwidth]{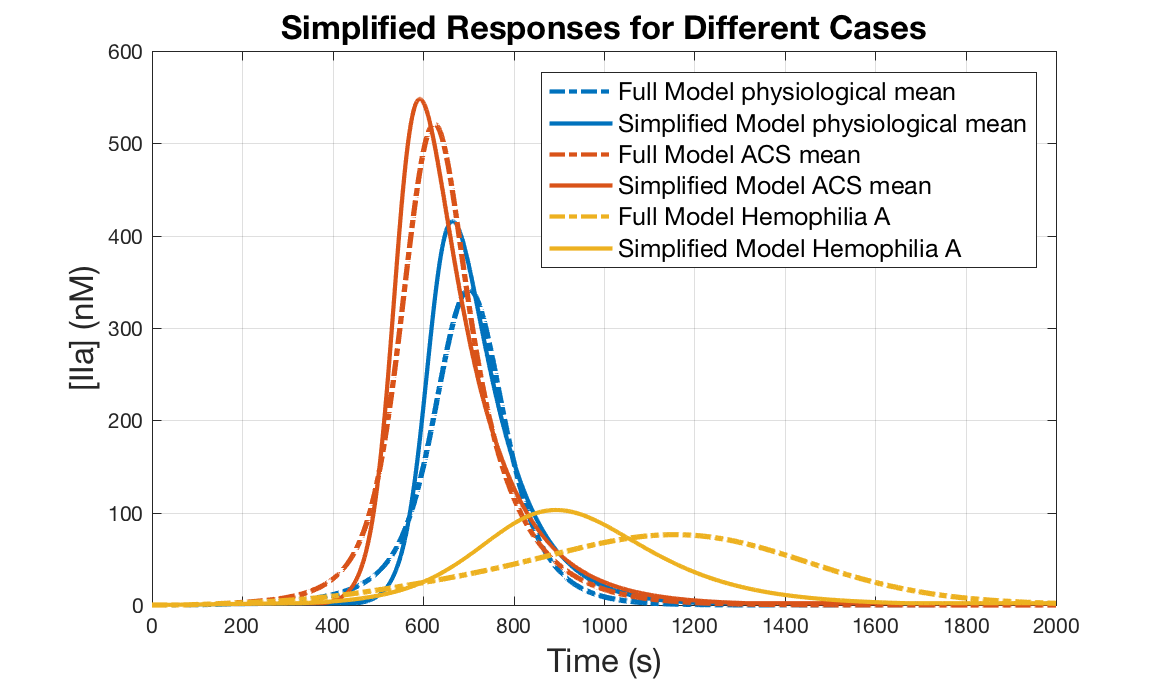}
\caption[Simplified model fit for different cases]{Simplified model fit for different cases. By changing the parameters, the model allows for a wide range of thrombin dynamics such as in acute coronary syndromes and hemophilia A.}  
\label{fig:simplifiedModel_3Cases}
\end{figure} 

Further, active thrombin could be propagated downstream and could potentially cause clotting elsewhere. This observation and model prediction on sustained activity of thrombin are hypothesized to play a necessary role towards effectively studying clotting in realistic geometries.

\subsection{Limitations and Improvements}

The simplified model proposed here, when coupled with fluid flow and transport models, is expected to make patient-specific mechanical characterization of clots in realistic geometries feasible. The parameters of the model need to be altered in order to better simulate thrombin dynamics on a wider range of plasma factor compositions. Figure \ref{fig:simplifiedModel_rates} shows simulations from parameters identified for acute coronary syndrome population mean composition \cite{brummel2008ACDvsCAD} and a hemophilia patient \cite{brummel2013riskdisease} (plasma factor composition of patient C in Figure 3. (B)). The hemophilia patient simulations suggest the form of rate functions as well as the threshold value (2 nM [IIa]) could be modeled better. 

\begin{figure}[htbp]
\centering
\includegraphics[width=0.6 \textwidth]{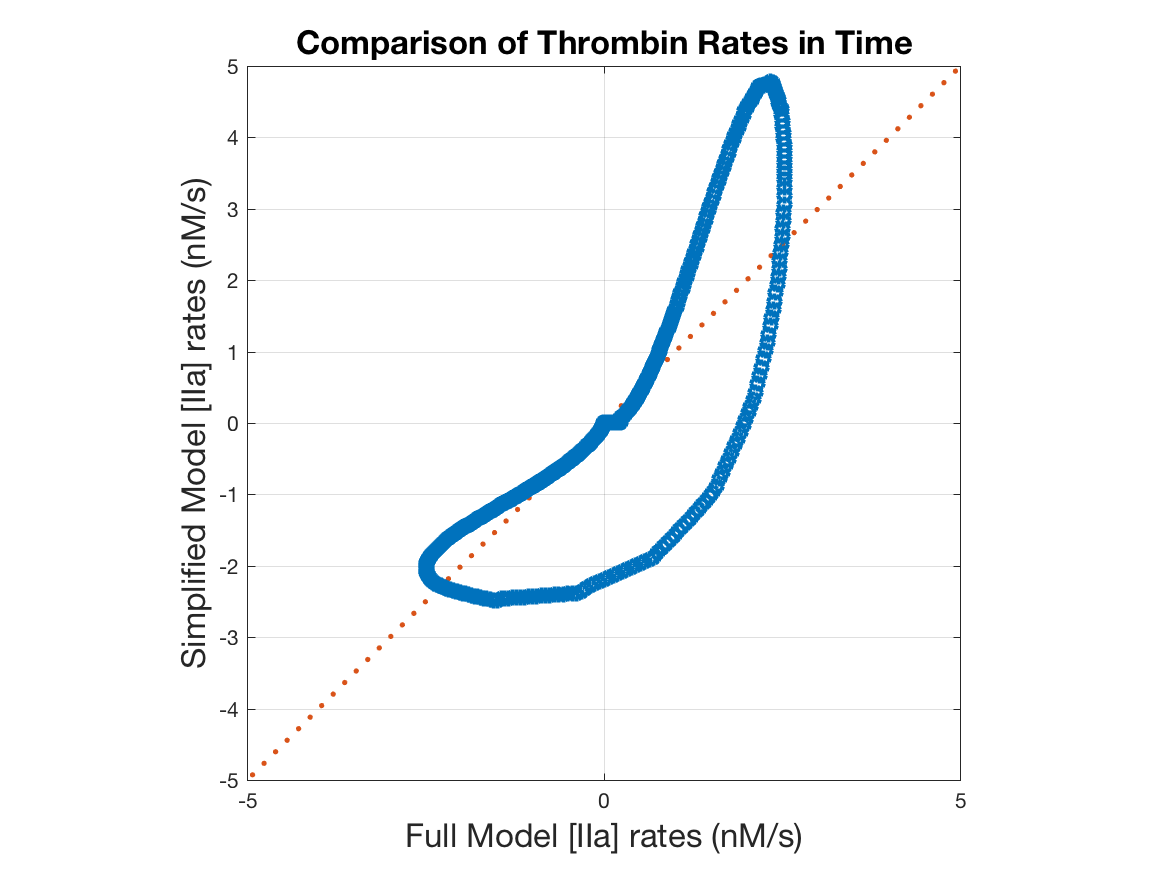}
\caption[Simplified model comparison of thrombin generation rates]{Comparison of thrombin generation rates in time. The better the data aligns with the $x=y$ line, the better is the simplification. The effect of time lag is better emphasized in this plot compared to Figure \ref{fig:ThrombinGen4SpeciesComparison}.}  
\label{fig:simplifiedModel_rates}
\end{figure} 

For the mean plasma composition, Figure \ref{fig:simplifiedModel_rates} shows the full model thrombin rates for the simplified model at different times during clotting plotted against the corresponding time points of the full model. The reaction rates could be improved to better align the simulation data with the $x=y$ line. We also note that the objective used for simplification in equation \ref{eq:simplificationObjective} does not explicitly penalize deviations in rates. Figure \ref{fig:simplifiedModel_rates} shows thrombin rates for the simplified model at different points in the state space along a particular trajectory of the full model plotted against the corresponding rates of the full model. The alignment is better in this case compared to figure \ref{fig:simplifiedModel_rates}. This suggests that the simplified model takes a slightly different path compared to the full model. We are currently working towards modifying the dynamics and the objective so that the rates could be better simulated. We are also currently testing the performance of the model while accounting for the effects of diffusion. 

\begin{figure}[htbp]
\centering
\includegraphics[width=0.6 \textwidth]{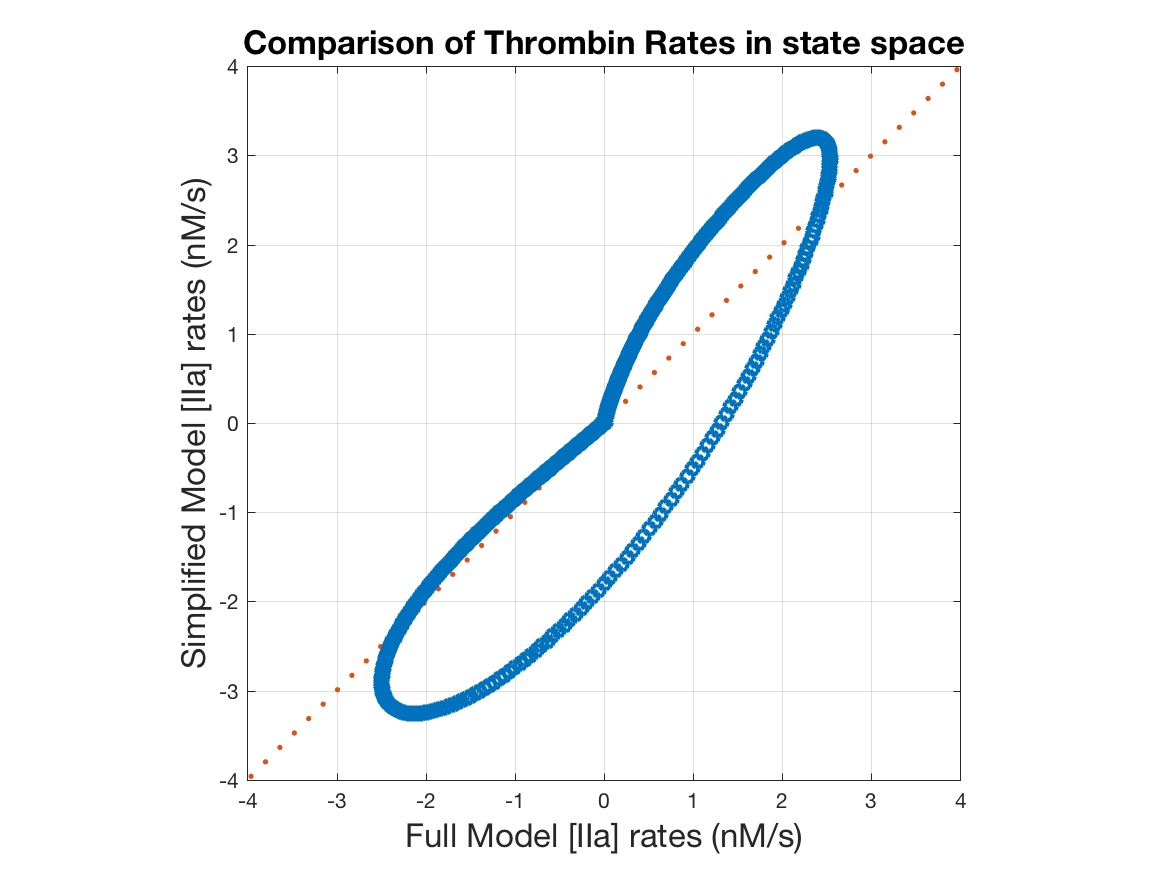}
\caption[Simplified model comparison of thrombin generation rates]{Comparison of thrombin generation rates along the mean physiological trajectory of the full model. The model predictions are better compared to those in Figure \ref{fig:simplifiedModel_rates}. }  
\label{fig:simplifiedModel_rates}
\end{figure} 

Augmenting the simplified model with fibrin dynamics would allow patient- and event-specific characterization of thromboelastography experiments \cite{mallett1992thrombelastography,mallett1993haemostasis}. This would be useful to better characterize clotting during surgeries \cite{kang1995thromboelastography}. This work could be further improved by coupling with simplified platelet aggregation models. To conclude, the drastically simplified model proposed here is a novel and a fertile step and subsequent progress has potential applications such as virtual flow-diverter treatment planning \cite{peach2016virtual}, predicting thrombotic risk accounting for  flow \cite{jordan2012flow}, factors like injuries in tandem \cite{jordan2011simulated} and stenosed arteries \cite{papadopoulos2016derivation}.



\section{Conclusion}
We proposed a simplified model for thrombin dynamics based on the stoichiometry of certain important chemical species. The model, using switching of parameters based on a threshold, describes dynamics of thrombin akin to jump starting a car. The simplified model fits corresponding aspects of the full model well and different features of thrombin dynamics are easily alterable. The model is able to predict certain important changes in thrombin dynamics due to changes in prothrombin and antithrombin concentration. This prediction using the simplified model with the potential for clinical manifestation in hypercoagulable diseases is hypothesized as a necessary step in patient-specific simulations of clotting in realistic geometries.







\bibliographystyle{unsrt}  
\bibliography{bibdataBlood_03BioMM2017,bibdataLearning_03BioMM2017}   


\end{document}